\documentclass[aps,prd,reprint,nofootinbib,longbibliography]{revtex4-1}

\usepackage{amsmath}
\usepackage{mathtools}
\usepackage{graphicx}
\usepackage[normalem]{ulem}
\usepackage[com pat=1.1.0]{tikz-feynman}
\usepackage{tikz}
\usepackage{aas_macros}
\usepackage{bm}
\usetikzlibrary{external}
\usepackage{comment}
\usetikzlibrary{arrows}
\usepackage[colorlinks=true, allcolors=blue]{hyperref}

\newcommand\barparena[1]{\overset{%
   \scriptscriptstyle(-)}{#1}}

\begin{document}

\title{Global features of fast neutrino-flavor conversion in binary neutron star merger}

\author{Hiroki Nagakura}
\email{hiroki.nagakura@nao.ac.jp}
\affiliation{Division of Science, National Astronomical Observatory of Japan, 2-21-1 Osawa, Mitaka, Tokyo 181-8588, Japan}

\begin{abstract}
Binary neutron star merger (BNSM) offers an environment where fast neutrino-flavor conversion (FFC) can vividly occur, that potentially leads to a considerable change of neutrino radiation field. In this paper, we investigate global features of FFC by general relativistic quantum kinetic neutrino transport simulations in spatial axisymmetry. Our result suggests that global advection of neutrinos plays a crucial role in FFC dynamics. Although flavor conversions occur ubiquitously in the early phase, they remain active only in a narrow region in the late phase. This region includes an ELN-XLN Zero Surface (EXZS), corresponding to a surface where electron-neutrinos lepton number (ELN) equals to heavy-leptonic one (XLN). The EXZS is not stationary, but dynamically evolve in a time scale of global advection. We also find that neutrinos can undergo a flavor swap when they pass through the EXZS, resulting in qualitatively different neutrino radiation fields between both sides of EXZS. Our result suggests that EXZS is one of the key ingredients to characterize FFC in BNSM.
\end{abstract}
\maketitle

\section{Introduction}\label{sec:intro}
Binary neutron star mergers (BNSM) are nowadays established as observable astrophysical transients through gravitational waves (GWs) \cite{2017PhRvL.119p1101A} and electromagnetic waves (EM) \cite{2017ApJ...848L..12A}. Though considerable progress in our understanding of BNSM has been achieved (see, e.g., recent reviews \cite{2017RPPh...80i6901B,2019ARNPS..69...41S,2020ARNPS..70...95R,2021ARA&A..59..155M}), neutrino quantum kinetics represents one of the most uncertain ingredients. Theoretical studies have suggested that neutrino flavor conversions ubiquitously occur in remnants of BNSM \cite{2016PhRvD..94j5006Z,2017PhRvD..95j3007W,2018PhRvD..97h3011V,2020PhRvD.102j3015G,2021PhRvL.126y1101L,2022PhRvD.106h3005R,2022PhRvD.105h3024J,2022arXiv220702214G,2022arXiv221203750X}. This implies that Boltzmann equation is not appropriate, and that quantum kinetic equation (QKE) governs the neutrino radiation field.

Fast neutrino-flavor conversion (FFC) \cite{2005PhRvD..72d5003S} potentially leads to a significant change of neutrino radiation field and subsequently affect fluid dynamics and nucleosynthesis (see, e.g., recent reviews \cite{2021ARNPS..71..165T,2022Univ....8...94C,2022arXiv220703561R,2023arXiv230111814V}). Unfortunately, however, the self-consistent simulation is still out of reach because of enormous computational demands. In fact, FFC can occur in the scale of sub-centimeters, which are several orders of magnitudes smaller than the global system. Previous studies are, hence, limited to the local simulations \cite{2022arXiv220702214G} or global simulations with a plane-parallel geometry and kilometer-scale spatial resolutions \cite{2021JCAP...01..017P}. 

Phenomenological approaches, in which the transport equation remains classical but flavor mixings are incorporated in a parametric manner, have also been made recently \cite{2021PhRvL.126y1101L,2022PhRvD.105h3024J,2022PhRvD.106j3003F}. It should be stressed that these results need to be considered provisional, since the results strongly depend on their treatments of flavor conversions. In fact, these models have parameters which control where, when, and how large flavor conversions occur. These ambiguities can be removed only by more physically accurate treatments by solving quantum kinetic neutrino transport. Such a first-principles approach also offers key information to develop better phenomenological models.

In this paper, we present global features of FFC in a remnant of BNSM by performing quantum kinetic neutrino transport simulations. This is the first-ever demonstration that QKE is solved in axisymmetry, while we employ a similar prescription used in our previous studies \cite{2022PhRvL.129z1101N,2023PhRvD.107f3033N,2023PhRvL.130u1401N} to make global simulations tractable. General relativistic (GR) and multi-dimensional effects are also taken into account. Our result suggests that non-linear evolution of FFC are qualitatively different from those obtained in local simulations. Unless otherwise noted, we work with $c = G = \hbar = 1$ unit, where $c$, $G$, and $\hbar$ are the light speed, the gravitational constant, and the reduced Planck constant, respectively.

\section{Methods and models}\label{sec:methomodel}
FFCs are essentially energy independent, unless energy-dependent neutrino-matter interactions are incorporated \cite{2022PhRvD.106j3029J,2022PhRvD.106l3013K,2023arXiv230316453K}; hence, we solve energy-integrated QKE. We also work in Schwarzschild spacetime with $M = 2.5 M_{\odot}$, where $M$ and $M_{\odot}$ denote a black hole mass and a solar mass, respectively, so as to take into account GR effects. Assuming axisymmetry in space and no collision term, the energy integrated QKE can be written as (see also \cite{2022PhRvD.106f3011N}),
\begin{equation}
  \begin{split}
& \frac{\partial }{\partial t} \biggl[ \Bigl(1 - \frac{2M}{r} \Bigr)^{-1/2} \barparena{\bm{f}} \biggr]
+ \frac{1}{r^2} \frac{\partial}{\partial r} \biggl[ r^2 \cos \theta_{\nu} \Bigl(1 - \frac{2M}{r} \Bigr)^{1/2}   \barparena{\bm{f}} \biggr] \\
& + \frac{1}{r \sin \theta} \frac{\partial}{\partial \theta} (  \sin \theta \sin \theta_{\nu} \cos \phi_{\nu} \barparena{\bm{f}}   ) \\
&- \frac{1}{\sin \theta_{\nu}} \frac{\partial}{\partial \theta_{\nu}} \biggl[ \sin^2 \theta_{\nu} \frac{r-3M}{r^2} \Bigl(1 - \frac{2M}{r} \Bigr)^{-1/2}   \barparena{\bm{f}} \biggl] \\
& - \frac{\cot \theta}{r}  \frac{\partial}{\partial \phi_{\nu}} ( \sin \theta_{\nu} \sin \phi_{\nu} \barparena{\bm{f}} )  =  - i \xi [\barparena{H},\barparena{\bm{f}}],
  \end{split}
\label{eq:SchQKE}
\end{equation}
where $\barparena{\bm{f}}$ denotes the energy-integrated density matrix of neutrinos, while the upper bar corresponds to that for the antineutrinos.
In this expression, $t$, $r$, $\theta$ represent time, radius, and zenith angle, respectively, while $\theta_{\nu}$ and $\phi_{\nu}$ denote polar- and azimuthal angles (measured from a radial coordinate basis) in momentum space. $\barparena{H}$ is the oscillation Hamiltonian that includes vacuum- and self-interaction potentials in this study. The vacuum potential is left with reduced mixing angles to trigger flavor conversions. Normal mass hierarchy is assumed, and we adopt squared mass differences of $\Delta m^2 = 2.5 \times 10^{-6} {\rm eV^2}$ for two flavor approximation, and $\Delta m^2_{21} = 7.42 \times 10^{-5} {\rm eV^2} $ and $\Delta m^2_{31} = 2.51 \times 10^{-3} {\rm eV^2}$ for three flavor framework. We set $10^{-6}$ for all mixing angles. Given the set of parameters, we evaluate the vacuum potential with $12$ MeV neutrino energy. $\xi (\le 1)$ denotes an attenuation parameter of oscillation Hamiltonian, which is a parameter to make the simulations tractable. We employ our GRQKNT code \cite{2022PhRvD.106f3011N} for all simulations.

\begin{figure}
\begin{minipage}{0.4\textwidth}
    \includegraphics[width=\linewidth]{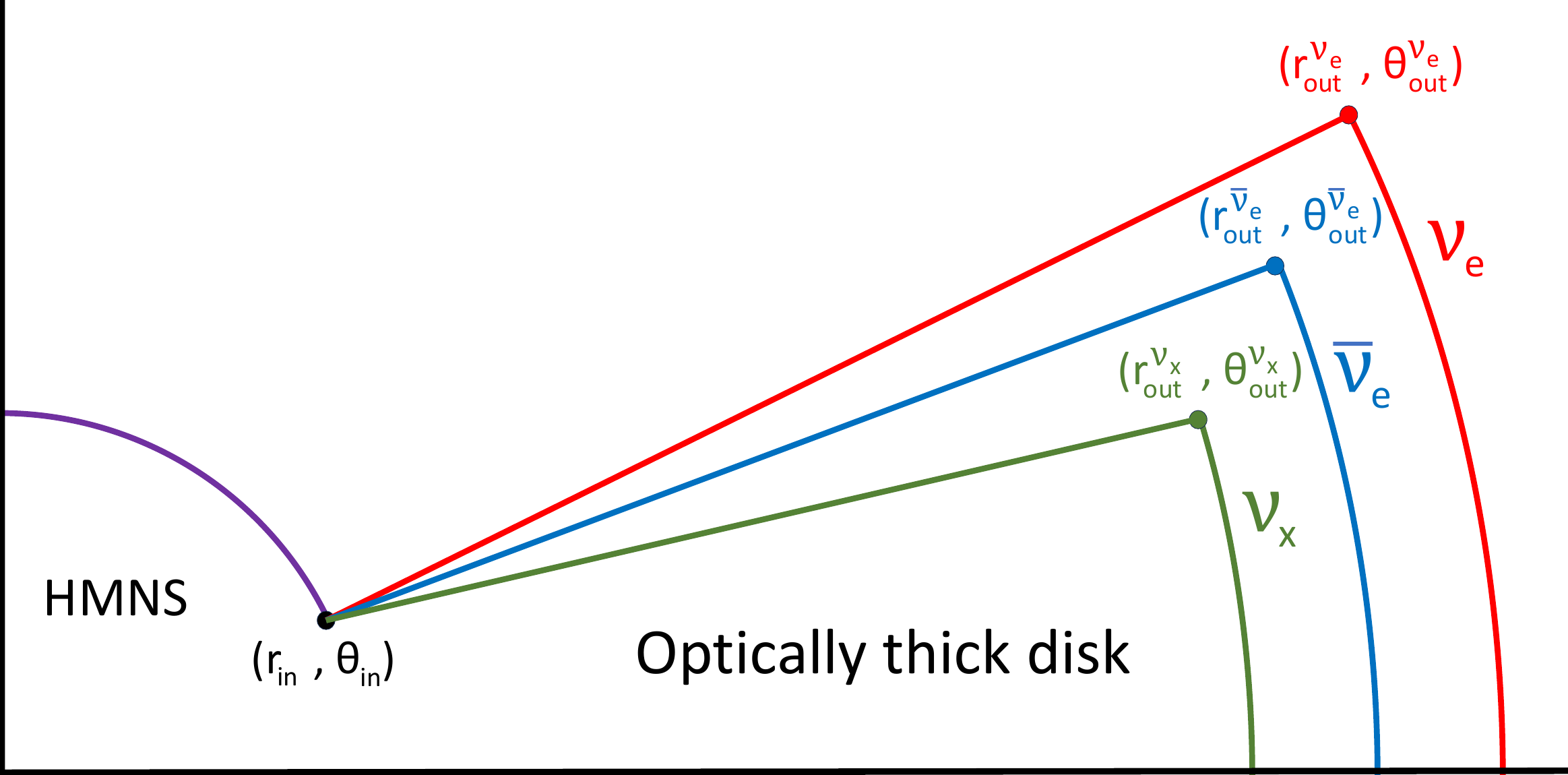}
\end{minipage}
    \caption{Schematic illustration of neutrino spheres. Neutrinos are radiated at each species-dependent neutrino sphere, which are distinguished by color: red, blue, and green for $\nu_e$, $\bar{\nu}_e$, and $\nu_x$, respectively. At the surface of hypermassive neutron star (HMNS) denoted by purple line, we assume all flavors of neutrinos are emitted.
}
    \label{Schematic}
\end{figure}

We assume that each flavor of neutrinos is emitted only at neutrino sphere. The schematic picture is provided in Fig.~\ref{Schematic}. Red, blue, and green solid lines represent neutrino spheres in the disk for electron-type neutrinos ($\nu_e$), their antipartners ($\bar{\nu}_e$), and other heavy leptonic neutrinos ($\nu_x$). Purple solid line represents the surface of hypermassive neutron star (HMNS) on which all flavor of neutrinos are radiated. Two remarks should be mentioned here. Although the HMNS is oblately deformed due to rotation in reality, it is assumed to be spherical; hence the geometry of the neutrino sphere can be determined by two parameters: $r_{\rm in}$ and $\theta_{\rm in}$. The former and the latter denote the radius of HMNS and the zenith angle where the disk is connected; we set $r_{\rm in}=15$km and $\theta_{\rm in}=60^{\circ}$ in this study. Second, the radius of neutrino sphere at the surface of HMNS is, in general, flavor dependent. However, the matter-density gradient at the surface of HMNS is very steep; consequently the difference of neutrino spheres is minor (see, e.g., \cite{2021ApJ...907...92S}). On the other hand, we set flavor dependent neutrino distributions in momentum space, in which we assume Fermi-Dirac distributions with zero chemical potentials, but temperature of neutrinos are different among flavors: $4, 4.5,$ and $5$ MeV for $\nu_e$, $\bar{\nu}_e$, and $\nu_x$, respectively. Since this setup may overestimate $\nu_x$ luminosities compared to realistic cases, its possible impacts on our results will also be discussed later.

The geometry of each neutrino sphere is characterized by $r^{\nu_\alpha}_{\rm out}$ and $\theta^{\nu_\alpha}_{\rm out}$, where $\alpha$ represents neutrino species (see Fig.~\ref{Schematic}). As a representative case, we set ($70$km, $60^{\circ}$), ($60$km, $65^{\circ}$), and ($55$km, $67^{\circ}$), for $\nu_e$, $\bar{\nu}_e$, and $\nu_x$, respectively. In the angular region from $\theta^{\nu_\alpha}_{\rm out}$ to $90^{\circ}$, the sphere is set on the radius of $r=r^{\nu_\alpha}_{\rm out}$. The equatorial symmetry is also assumed. In all simulations, we cover from $15$km to $100$km. The energy spectra of neutrinos are the same as those set on the surface of HMNS.

\begin{figure}
\begin{minipage}{0.5\textwidth}
    \includegraphics[width=\linewidth]{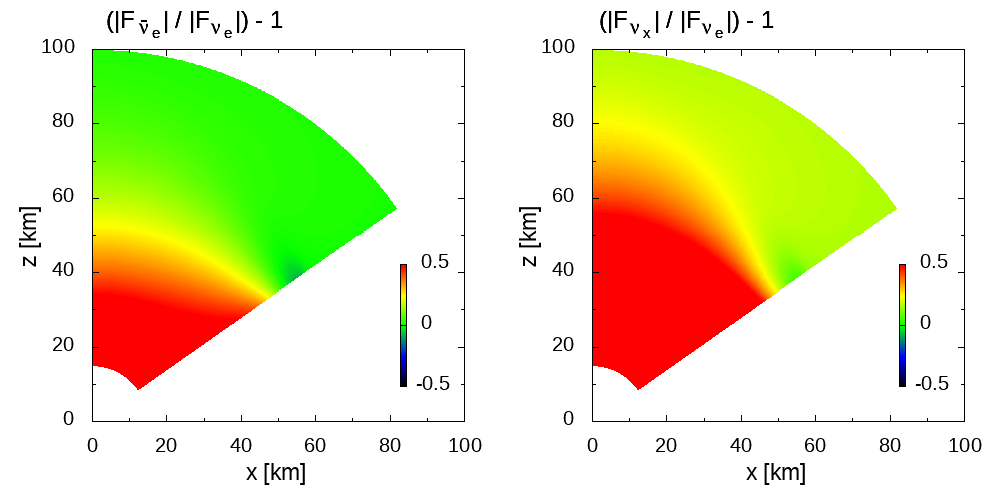}
\end{minipage}
    \caption{The deviation of number flux ratios from $\nu_e$ for a steady-state radiation field obtained from simulations with $H=0$ (Mno). In the left panel, we show $(|F_{\bar{\nu}_e}|/|F_{\nu_e}|)-1$, while $(|F_{\nu_x}|/|F_{\nu_e}|)-1$ is displayed for the right one.
}
    \label{graph_Flux_ratioNue}
\end{figure}

Before carrying out QKE simulations, we run a simulation with $H=0$ in Eq.~\ref{eq:SchQKE} (hereafter referred to as Mno model). This simulation is stopped at the time when the system has already reached a steady state ($1$ms). The steady state profile is also used as an initial condition for QKE simulations. In Fig.~\ref{graph_Flux_ratioNue}, we provide information on neutrino fluxes in the steady state. In the inner region ($\lesssim 50 {\rm km}$), both $\bar{\nu}_e$ and $\nu_x$ fluxes are remarkably higher than $\nu_e$, which is mainly attributed to the fact that they have higher temperature than $\nu_e$ at each neutrino sphere. At large radii, however, the difference of number flux among different flavors tends to be reduced, which is mainly due to geometrical effects of neutrino spheres (see Fig.~\ref{Schematic}). In fact, the larger surface area of neutrino sphere for $\nu_e$ than others can compensate for the lower temperature. It is also worth noting that the $\nu_e$ flux can overwhelm $\bar{\nu}_e$ and $\nu_x$ in the vicinity of $\nu_e$ neutrino sphere with $r^{\bar{\nu}_e}_{\rm out} \lesssim r \lesssim r^{\nu_e}_{\rm out}$ and $r^{\nu_x}_{\rm out} \lesssim r \lesssim r^{\nu_e}_{\rm out}$, respectively. We also note that electron-neutrinos lepton number (ELN) crossings are ubiquitously observed above the disk, which is in line with the result in \cite{2017PhRvD..95j3007W}.

\begin{figure*}
\begin{minipage}{1.0\textwidth}
    \includegraphics[width=\linewidth]{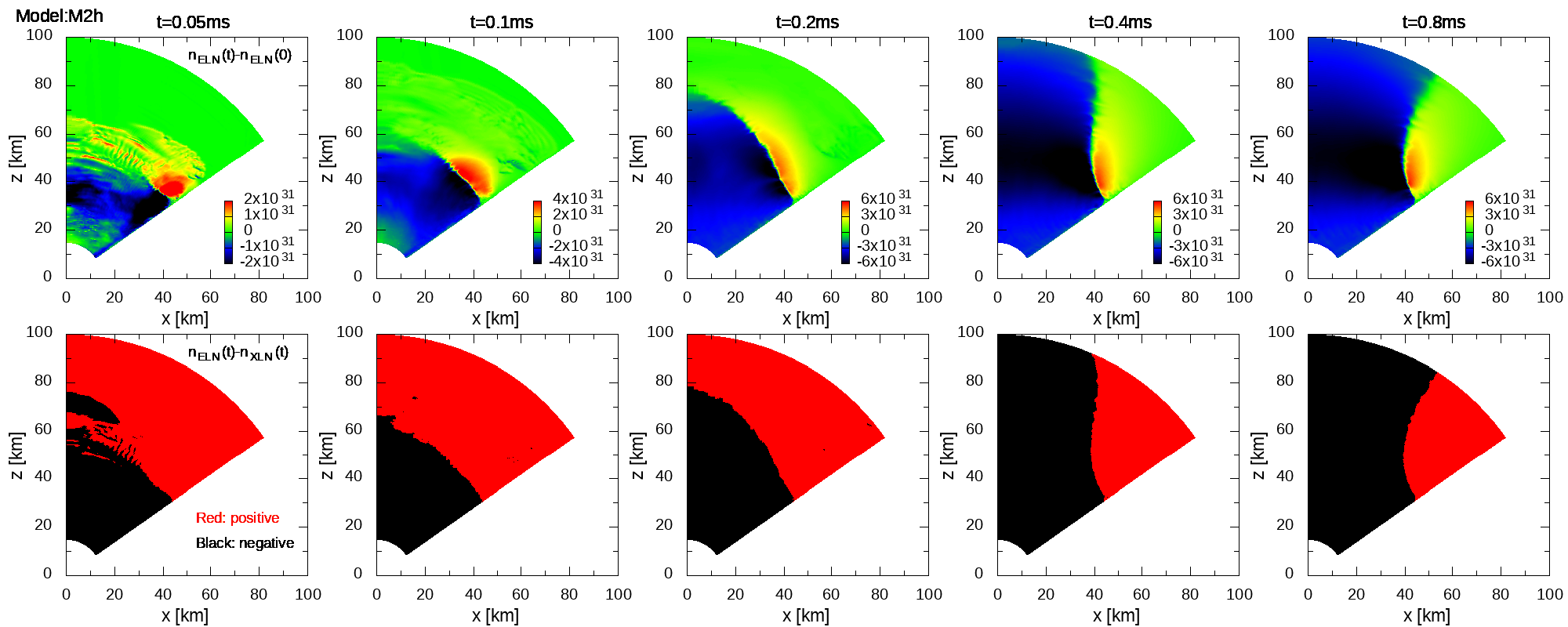}
\end{minipage}
    \caption{Top: difference of ELN number density from the initial distribution, $n_{\rm ELN} (t)-n_{\rm ELN} (0)$. Bottom: ELN-XLN number density, $n_{\rm ELN-XLN}(t)$, in which we highlight its sign by red- (for positive) and black (for negative) colors. The boundary between the two regions is a ELN-XLN Zero Surface (EXZS); see the text for more details. We displayed them at five different time snapshots for M2h: $t=0.05, 0.1, 0.2, 0.4,$ and $0.8$ ms from left to right. At $t=0$ms, the EXZS is located at $\sim 80$ km along the pole and it smoothly connects to $\sim 40$ km at $\theta = 55^{\circ}$.
}
    \label{graph_tdepeCmap2D_ELN_XLN_Numden_2flav}
\end{figure*}

In QKE simulations, we only focus on the spatial region above the disk; the simulations are conducted in the region of $0^{\circ} \le \theta \le 55^{\circ}$. The boundary at $\theta = 55^{\circ}$ is located slightly above the neutrino spheres, which allows us to easily handle the boundary condition for neutrinos. We use Dirichlet boundary conditions (constant in time) for neutrinos which come into computational domain, while we adopt a zero-gradient free boundary condition for neutrinos going out from the simulation box. A reflective boundary condition is adopted along the z-axis ($\theta=0^{\circ}$). We consider two models in this study: M3h and M2h. They have the same numerical setup except that the former and the latter corresponds to three- and two-flavor frameworks, respectively. 

We adopt a non-uniform radial grid (see \cite{2022PhRvL.129z1101N,2023PhRvD.107f3033N,2023PhRvL.130u1401N}) with $\Delta r_{\rm min}=3$m, where $\Delta r_{\rm min}$ denotes the radial width of the innermost mesh. The number of radial cells is 1152. In the meridian direction, we set a uniform grid for the cosine of the zenith-angle. The number of grid points is 384. Neutrino angles in momentum space are covered by a uniform grid with respect to $\cos \theta_{\nu}$ and $\phi_{\nu}$ with $96 \times 48$ grid points. We run the simulations up to $t=0.8$ms. $\xi$ is set to be $3 \times 10^{-4}$. Although qualitative trends are captured in these simulations (we made a resolution study), the reduction of $\xi$ causes some artificial results. This possible impact will also be discussed later.

\section{Results}\label{sec:reso}
Soon after QKE simulations begin, FFC occurs vigorously in the vicinity of HMNS and the disk. During the very early phase, FFC dynamics is characterized only by local properties of neutrino radiation field. As discussed in \cite{2023PhRvD.107f3033N,2023arXiv230405044Z}, however, non-linear states of FFC are substantially changed in the time scale of global advection. This is illustrated in the top panels of Fig.~\ref{graph_tdepeCmap2D_ELN_XLN_Numden_2flav}, in which we show the spatial distribution of $n_{\rm ELN}$ (ELN number density) subtracted by that of the initial one for M2h model. One noticeable feature is that the positive and negative region of $n_{\rm ELN}(t)-n_{\rm ELN}(0)$ is clearly separated from each other by a very narrow region at $t \gtrsim 0.1$ms. We also find that that the transition region includes a line (or surface if we take into account the azimuthal direction) where the sign of the number density of ELN-XLN ($n_{\rm ELN-XLN}(t) \equiv n_{\rm ELN}(t)-n_{\rm XLN}(t)$) is flipped (see bottom panels of Fig.~\ref{graph_tdepeCmap2D_ELN_XLN_Numden_2flav}). Here, XLN denotes a heavy-leptonic-neutrinos lepton number. We call the surface as ELN-XLN Zero Surface (EXZS).

\begin{figure}
\begin{minipage}{0.35\textwidth}
    \includegraphics[width=\linewidth]{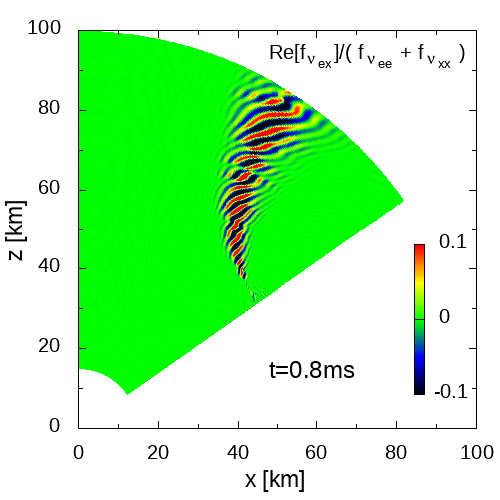}
\end{minipage}
    \caption{Color map for the ratio of the real part of the off-diagonal component of $\bm{f}$ to the sum of the diagonal elements at $t=0.8$ms for M2h model.
}
    \label{graph_OffD}
\end{figure}

As shown in Fig.~\ref{graph_tdepeCmap2D_ELN_XLN_Numden_2flav}, the EXZS at the $\theta$-boundary ($\theta_{\rm max}$) does not evolve with time. This is attributed to the geometry of neutrino spheres. In our models, $\bar{\nu}_e$ is more populated than $\nu_e$ in the inner region ($r \lesssim r^{\bar{\nu}_e}_{\rm out}$). On the other hand, the number density of $\nu_e$ can dominate over $\bar{\nu}_e$ at $r \sim r^{\nu_e}_{\rm out}$ and $\theta \sim \theta^{\nu_e}_{\rm out}$. This indicates that EXZS appears at $t=0$ms (we note $n_{\rm XLN}(0)=0$). Since the neutrinos injected into computational domain from $\theta = \theta_{\rm max}$ are assumed to be constant in time, the EXZS is nearly fixed.

\begin{figure}
\begin{minipage}{0.5\textwidth}
    \includegraphics[width=\linewidth]{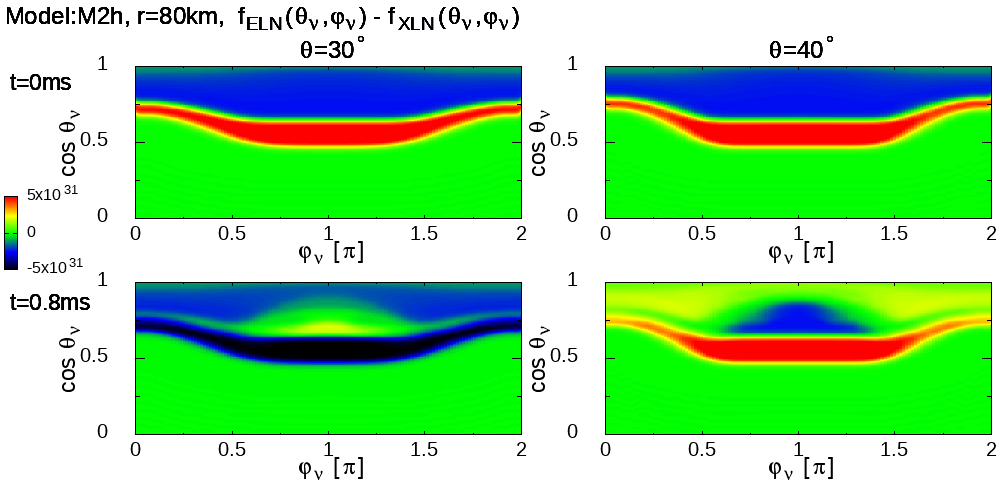}
\end{minipage}
    \caption{ELN-XLN angular distributions for M2h. We focus on the region of $\cos {\theta}_{\nu} >0$ (outgoing neutrinos in the radial direction). In top and bottom panels, we show the result at $t=0$ and $0.8$ms, respectively. The left and right panels distinguish the spatial position: $\theta=30^{\circ}$ and $40^{\circ}$ at $r=80$km.
}
    \label{graph_angdistri_ELN_XLN}
\end{figure}

However, the EXZS in the region of $\theta < \theta_{\rm max}$ is not stationary. As shown in the bottom panels of Fig.~\ref{graph_tdepeCmap2D_ELN_XLN_Numden_2flav}, the negative region of $n_{\rm ELN-XLN}$ expands with time, in particular around the z-axis. It seems that the expansion velocity of EXZS is associated with a neutrino flux balance between both sides, whose detailed properties will be investigated in our forthcoming paper, though. We find that $\nu_e$ in the angular region where $\bar{\nu}_e$ is absent (see also Fig.~1 in \cite{2017PhRvD..95j3007W}) converts into $\nu_x$, that is mainly responsible for the negative $n_{\rm ELN-XLN}$ there. We also find that the flavor conversion is less vigorous around the z-axis, despite the fact that the neutrino radiation field is substantially changed. Indeed, the amplitude of off-diagonal components of $\bm{f}$ is less than a percent of the diagonal ones around the pole. On the other hand, it is $>10 \%$ around the EXZS at $t=0.8$ms (see Fig.~\ref{graph_OffD}), suggesting that flavor conversions occur only in the vicinity of EXZS, and then the FFC-experienced neutrinos advect to other regions.

It may be interesting to compare the EXZS to {\it streamlines} of neutrinos discussed in \cite{2008PhRvD..78c3014D}. Each streamline portrays the trajectory of total neutrino flux, defined from the trace parts of density matrix of neutrinos, i.e., representing the overall neutrino advection in space. One thing we do notice here is, however, that EXZS is not directly related to them; the reason is as follows. As is well known, the trace part of density matrix is unaffected by flavor conversions, implying that the total neutrino flux remains constant in time. On the other hand, the EXZS is dynamically evolved as shown in Fig.~\ref{graph_tdepeCmap2D_ELN_XLN_Numden_2flav}, indicating that it is not dictated by the total neutrino flux. This argument is also supported by the definition of EXZS, which is not characterized by the trace part of density matrix but rather ELN-XLN.

We compare ELN-XLN angular distributions between two different spatial positions in Fig.~\ref{graph_angdistri_ELN_XLN}. The left and right panels portray distributions at $\theta=30^{\circ}$ and $40^{\circ}$, respectively, on the same radius ($r=80$km). We note that the EXZS is located between the two positions. At $t=0$ms (top panels), the angular distributions of neutrinos at the two spatial positions are nearly identical to each other. At $\cos {\theta}_{\nu} \sim 1$, ELN is negative, that is due to the higher emission of $\bar{\nu}_e$ than $\nu_e$ at the neutrino sphere. However, there is a region (band) where $\nu_e$ dominates over $\bar{\nu}_e$ around $\cos {\theta}_{\nu} \sim 0.6$. In this region, $\bar{\nu}_e$ is absent due to the smaller size of neutrino sphere in the disk than that of $\nu_e$ (see Fig.~\ref{Schematic}). At $t=0.8$ms, however, the ELN-XLN in the band becomes negative at the $\theta=30^{\circ}$ position, suggesting that these neutrinos undergo an almost complete swap from $\nu_e$ to $\nu_x$. As we mentioned above, the flavor swap occurs around EXZS and then advect to the spatial position. On the other hand, the neutrinos in $\cos {\theta}_{\nu} \sim 1$ is almost the same at $t=0$ms, suggesting that FFC is inefficient for these neutrinos. 

ELN-XLN angular distributions at $\theta=40^{\circ}$ displayed in the right bottom panel of Fig.~\ref{graph_angdistri_ELN_XLN} have an opposite trend to the case with $\theta=30^{\circ}$. Neutrinos in $\cos {\theta}_{\nu} \sim 1$, which has already passed through the EXZS, experienced strong flavor conversions, whereas those in the angular band at $\cos {\theta}_{\nu} \sim 0.6$ remain $\nu_e$-dominant. This indicates that the change of neutrino radiation field can be quantified by how many neutrinos pass through the EXZS.

\begin{figure}
\begin{minipage}{0.48\textwidth}
    \includegraphics[width=\linewidth]{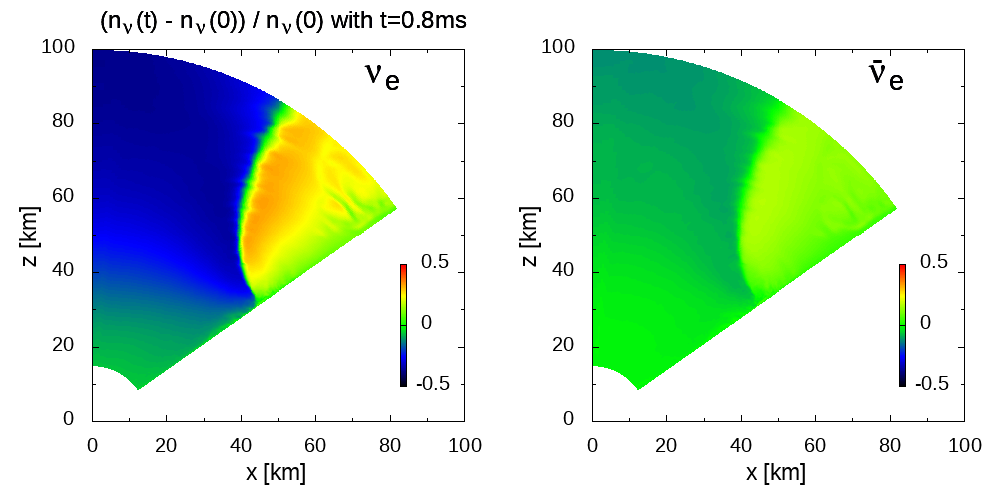}
\end{minipage}
    \caption{The difference of neutrino number density between $t=0.8$ and $0$ms, which is normalized by the density at $t=0$ms. Left: $\nu_e$. Right: $\bar{\nu}_e$. The model is M3h.
}
    \label{graph_Nue_Nueb_diffromini}
\end{figure}

We should remark a caveat here. As shown in the bottom panels of Fig.~\ref{graph_angdistri_ELN_XLN}, there are ELN-XLN angular crossings, albeit shallow, indicating that neutrinos are still in unstable state (see also \cite{2022arXiv221101398N,2023arXiv230405044Z}). The suppression of FFC may be an artifact due to the reduction of oscillation Hamiltonian by $\xi$. This issue can be addressed by increasing $\xi$, albeit requiring higher spatial resolutions.

Finally, we quantify how large $\nu_e$ and $\bar{\nu}_e$ radiation fields are changed due to FFCs. As displayed in Fig.~\ref{graph_Nue_Nueb_diffromini}, both $\nu_e$ and $\bar{\nu}_e$ number densities in M3h model become lower than Mno in the polar region. The change of $n_{\nu_e}$ reaches $\sim 50 \%$ at $r \sim 100$km. $\bar{\nu}_e$ is a bit moderate change, which is because the difference of $\bar{\nu}_e$- and $\bar{\nu}_x$ spheres and their energy spectra are smaller than those for $\nu_e$ and $\nu_x$. The flavor swap ($\barparena{\nu}_e$ conversion into $\barparena{\nu}_x$) at EXZS accounts for the reduction of $\barparena{\nu}_e$ around the polar region, as discussed above.

On the other hand, both $n_{\nu_e}$ and $n_{\bar{\nu}_e}$ become higher than those at $t=0$ms in the region of $r \gtrsim 50$km and close to the disk. As we have already mentioned, the neutrinos in $\cos {\theta}_{\nu} \sim 1$ has an experience of FFC when they pass through the EXZS. Since $\barparena{\nu}_x$ dominates over $\barparena{\nu}_e$ in the angular region at $t=0$ms, FFC facilitates the increase of $\barparena{\nu}_e$. 

\begin{figure}
\begin{minipage}{0.48\textwidth}
    \includegraphics[width=\linewidth]{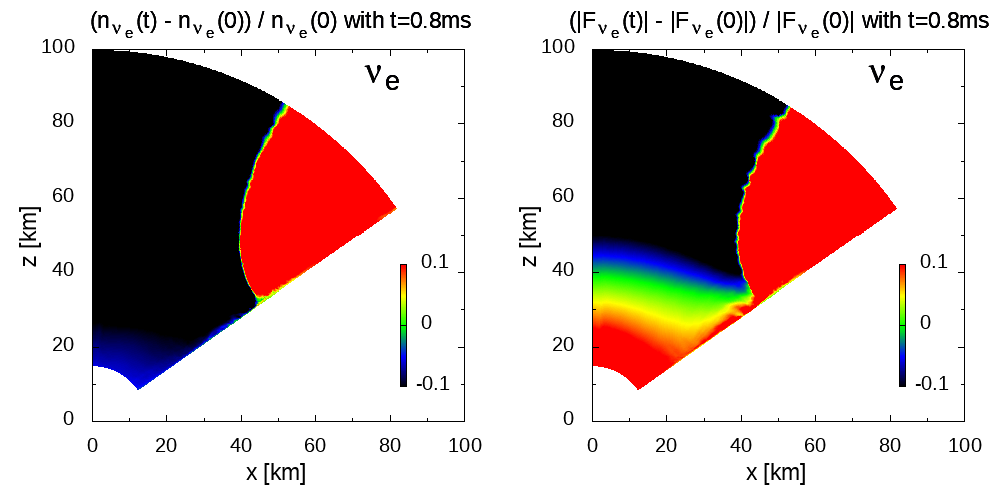}
\end{minipage}
    \caption{Similar as Fig.~\ref{graph_Nue_Nueb_diffromini} but for comparisons between the number density and flux, by focusing on $\nu_e$ in M3h model. The left panel corresponds to the same one in Fig.~\ref{graph_Nue_Nueb_diffromini}, but the range of color contour is different. The right panel displays the result for the norm of neutrino flux.
}
    \label{graph_AbsFlux_Num_nue_diffromini}
\end{figure}

The number flux of $\barparena{\nu}_e$ also has the similar trend as the number density if FFCs occur (see Fig.~\ref{graph_AbsFlux_Num_nue_diffromini}), but the noticeable difference emerges around the neutrino sphere; the norm of flux becomes higher at $t=0.8$ms than $t=0$ms there. This is attributed to the fact that the total number of radially-incoming $\nu_e$, emitted from the neutrino sphere at large radii, becomes lower due to FFCs. This leads to the reduction of the number density of $\nu_e$ (as discussed above) but increases the anisotropy of $\nu_e$ in the region. The latter effect dominates over the former, resulting in increasing the norm of $\nu_e$ flux.

An important caveat needs to be mentioned here. The above conclusion regarding the impact of FFCs on $\barparena{\nu}_e$ strongly hinges on $\nu_x$ radiation field. After FFC occurs, $\barparena{\nu}_e$ may be decreased if $\barparena{\nu}_x$ number density is smaller than that assumed in Mno model. It is, hence, important to develop accurate models of $\nu_x$ radiation field in order to draw a robust conclusion of the actual impact of FFCs on BNSM. It should also be noted that the disappearance of $\nu_x$ sphere, which is usually accompanied by black hole formation, results in a substantial reduction of $\barparena{\nu}_e$ if FFCs occur. Such environments seem to emerge, albeit temporally, in BNSM remnant systems. It is an intriguing question how FFCs can impact fluid dynamics and nucleosynthesis. We deffer the detailed investigations for these intriguing issues to future work.

\section{Conclusions}\label{sec:conclu}
In this paper, we discuss global features of FFC in a geometry of BNSM remnant with HMNS, based on QKE simulations with attenuating the oscillation Hamiltonian. Our result suggests that effects of global advection lead to a substantial change of FFC dynamics. The most striking result in this study is that FFC properties are qualitatively different between the regions divided by a ELN-XLN Zero Surface (EXZS). The surface evolves with time due to global neutrino advection. When neutrinos pass through the EXZS, flavor swaps occur in the very narrow region. It should also be mentioned that whether neutrinos pass through the EXZS or not hinges on neutrino trajectories, indicating that occurrences of flavor swaps depend on neutrino flight directions. This fact also exhibits that a complex neutrino radiation field emerges, depending on the time-dependent and global structure of EXZS.

Although we show that the EXZS is a key ingredient to characterize FFC in BNSM system, there is very little known about its property. We also note that FFC dynamics hinges on species-dependent neutrino energy spectrum and the geometry of neutrino spheres. Since both HMNS and the disk evolve with time, a systematic study with varying these setups is necessary to make more definitive claims about roles of FFC on BNSM dynamics. Neglecting neutrino-matter interactions is also another concern in the present study. In fact, multiple EXZSs may emerge in the complex fluid distributions. Addressing these issues is the next goal of our future research.

\section{Acknowledgments}
We are grateful to Masamichi Zaizen, Lucas Johns, Chinami Kato, and Kohsuke Sumiyoshi for useful comments and discussions. This work is supported by the HPCI System Research Project (Project ID: 220173, 220047, 220223, 230033, 230204, 230270), XC50 of CfCA at the National Astronomical Observatory of Japan (NAOJ), Yukawa-21 at Yukawa Institute for Theoretical Physics of Kyoto University, and the High Energy Accelerator Research Organization (KEK). For providing high performance computing resources, Computing Research Center, KEK, and JLDG on SINET of NII are acknowledged. HN is supported by Grant-inAid for Scientific Research (23K03468).
\bibliography{bibfile}

\begin{thebibliography}{33}%
\makeatletter
\providecommand \@ifxundefined [1]{%
 \@ifx{#1\undefined}
}%
\providecommand \@ifnum [1]{%
 \ifnum #1\expandafter \@firstoftwo
 \else \expandafter \@secondoftwo
 \fi
}%
\providecommand \@ifx [1]{%
 \ifx #1\expandafter \@firstoftwo
 \else \expandafter \@secondoftwo
 \fi
}%
\providecommand \natexlab [1]{#1}%
\providecommand \enquote  [1]{``#1''}%
\providecommand \bibnamefont  [1]{#1}%
\providecommand \bibfnamefont [1]{#1}%
\providecommand \citenamefont [1]{#1}%
\providecommand \href@noop [0]{\@secondoftwo}%
\providecommand \href [0]{\begingroup \@sanitize@url \@href}%
\providecommand \@href[1]{\@@startlink{#1}\@@href}%
\providecommand \@@href[1]{\endgroup#1\@@endlink}%
\providecommand \@sanitize@url [0]{\catcode `\\12\catcode `\$12\catcode
  `\&12\catcode `\#12\catcode `\^12\catcode `\_12\catcode `\%12\relax}%
\providecommand \@@startlink[1]{}%
\providecommand \@@endlink[0]{}%
\providecommand \url  [0]{\begingroup\@sanitize@url \@url }%
\providecommand \@url [1]{\endgroup\@href {#1}{\urlprefix }}%
\providecommand \urlprefix  [0]{URL }%
\providecommand \Eprint [0]{\href }%
\providecommand \doibase [0]{http://dx.doi.org/}%
\providecommand \selectlanguage [0]{\@gobble}%
\providecommand \bibinfo  [0]{\@secondoftwo}%
\providecommand \bibfield  [0]{\@secondoftwo}%
\providecommand \translation [1]{[#1]}%
\providecommand \BibitemOpen [0]{}%
\providecommand \bibitemStop [0]{}%
\providecommand \bibitemNoStop [0]{.\EOS\space}%
\providecommand \EOS [0]{\spacefactor3000\relax}%
\providecommand \BibitemShut  [1]{\csname bibitem#1\endcsname}%
\let\auto@bib@innerbib\@empty
\bibitem [{\citenamefont {{Abbott}}\ \emph
  {et~al.}(2017{\natexlab{a}})\citenamefont {{Abbott}}, \citenamefont
  {{Abbott}}, \citenamefont {{Abbott}}, \citenamefont {{Acernese}},
  \citenamefont {{Ackley}}, \citenamefont {{Adams}}, \citenamefont {{Adams}},
  \citenamefont {{Addesso}}, \citenamefont {{Adhikari}}, \citenamefont
  {{Adya}},\ and\ \citenamefont {et~al.}}]{2017PhRvL.119p1101A}%
  \BibitemOpen
  \bibfield  {author} {\bibinfo {author} {\bibfnamefont {B.~P.}\ \bibnamefont
  {{Abbott}}}, \bibinfo {author} {\bibfnamefont {R.}~\bibnamefont {{Abbott}}},
  \bibinfo {author} {\bibfnamefont {T.~D.}\ \bibnamefont {{Abbott}}}, \bibinfo
  {author} {\bibfnamefont {F.}~\bibnamefont {{Acernese}}}, \bibinfo {author}
  {\bibfnamefont {K.}~\bibnamefont {{Ackley}}}, \bibinfo {author}
  {\bibfnamefont {C.}~\bibnamefont {{Adams}}}, \bibinfo {author} {\bibfnamefont
  {T.}~\bibnamefont {{Adams}}}, \bibinfo {author} {\bibfnamefont
  {P.}~\bibnamefont {{Addesso}}}, \bibinfo {author} {\bibfnamefont {R.~X.}\
  \bibnamefont {{Adhikari}}}, \bibinfo {author} {\bibfnamefont {V.~B.}\
  \bibnamefont {{Adya}}}, \ and\ \bibinfo {author} {\bibnamefont {et~al.}},\
  }\bibfield  {title} {\enquote {\bibinfo {title} {{GW170817: Observation of
  Gravitational Waves from a Binary Neutron Star Inspiral}},}\ }\href {\doibase
  10.1103/PhysRevLett.119.161101} {\bibfield  {journal} {\bibinfo  {journal}
  {Physical Review Letters}\ }\textbf {\bibinfo {volume} {119}},\ \bibinfo
  {eid} {161101} (\bibinfo {year} {2017}{\natexlab{a}})},\ \Eprint
  {http://arxiv.org/abs/1710.05832} {arXiv:1710.05832 [gr-qc]} \BibitemShut
  {NoStop}%
\bibitem [{\citenamefont {{Abbott}}\ \emph
  {et~al.}(2017{\natexlab{b}})\citenamefont {{Abbott}}, \citenamefont
  {{Abbott}}, \citenamefont {{Abbott}}, \citenamefont {{Acernese}},
  \citenamefont {{Ackley}}, \citenamefont {{Adams}}, \citenamefont {{Adams}},
  \citenamefont {{Addesso}}, \citenamefont {{Adhikari}}, \citenamefont
  {{Adya}},\ and\ \citenamefont {et~al.}}]{2017ApJ...848L..12A}%
  \BibitemOpen
  \bibfield  {author} {\bibinfo {author} {\bibfnamefont {B.~P.}\ \bibnamefont
  {{Abbott}}}, \bibinfo {author} {\bibfnamefont {R.}~\bibnamefont {{Abbott}}},
  \bibinfo {author} {\bibfnamefont {T.~D.}\ \bibnamefont {{Abbott}}}, \bibinfo
  {author} {\bibfnamefont {F.}~\bibnamefont {{Acernese}}}, \bibinfo {author}
  {\bibfnamefont {K.}~\bibnamefont {{Ackley}}}, \bibinfo {author}
  {\bibfnamefont {C.}~\bibnamefont {{Adams}}}, \bibinfo {author} {\bibfnamefont
  {T.}~\bibnamefont {{Adams}}}, \bibinfo {author} {\bibfnamefont
  {P.}~\bibnamefont {{Addesso}}}, \bibinfo {author} {\bibfnamefont {R.~X.}\
  \bibnamefont {{Adhikari}}}, \bibinfo {author} {\bibfnamefont {V.~B.}\
  \bibnamefont {{Adya}}}, \ and\ \bibinfo {author} {\bibnamefont {et~al.}},\
  }\bibfield  {title} {\enquote {\bibinfo {title} {{Multi-messenger
  Observations of a Binary Neutron Star Merger}},}\ }\href {\doibase
  10.3847/2041-8213/aa91c9} {\bibfield  {journal} {\bibinfo  {journal} {\apjl}\
  }\textbf {\bibinfo {volume} {848}},\ \bibinfo {eid} {L12} (\bibinfo {year}
  {2017}{\natexlab{b}})},\ \Eprint {http://arxiv.org/abs/1710.05833}
  {arXiv:1710.05833 [astro-ph.HE]} \BibitemShut {NoStop}%
\bibitem [{\citenamefont {{Baiotti}}\ and\ \citenamefont
  {{Rezzolla}}(2017)}]{2017RPPh...80i6901B}%
  \BibitemOpen
  \bibfield  {author} {\bibinfo {author} {\bibfnamefont {Luca}\ \bibnamefont
  {{Baiotti}}}\ and\ \bibinfo {author} {\bibfnamefont {Luciano}\ \bibnamefont
  {{Rezzolla}}},\ }\bibfield  {title} {\enquote {\bibinfo {title} {{Binary
  neutron star mergers: a review of Einstein{\textquoteright}s richest
  laboratory}},}\ }\href {\doibase 10.1088/1361-6633/aa67bb} {\bibfield
  {journal} {\bibinfo  {journal} {Reports on Progress in Physics}\ }\textbf
  {\bibinfo {volume} {80}},\ \bibinfo {eid} {096901} (\bibinfo {year}
  {2017})},\ \Eprint {http://arxiv.org/abs/1607.03540} {arXiv:1607.03540
  [gr-qc]} \BibitemShut {NoStop}%
\bibitem [{\citenamefont {{Shibata}}\ and\ \citenamefont
  {{Hotokezaka}}(2019)}]{2019ARNPS..69...41S}%
  \BibitemOpen
  \bibfield  {author} {\bibinfo {author} {\bibfnamefont {Masaru}\ \bibnamefont
  {{Shibata}}}\ and\ \bibinfo {author} {\bibfnamefont {Kenta}\ \bibnamefont
  {{Hotokezaka}}},\ }\bibfield  {title} {\enquote {\bibinfo {title} {{Merger
  and Mass Ejection of Neutron Star Binaries}},}\ }\href {\doibase
  10.1146/annurev-nucl-101918-023625} {\bibfield  {journal} {\bibinfo
  {journal} {Annual Review of Nuclear and Particle Science}\ }\textbf {\bibinfo
  {volume} {69}},\ \bibinfo {pages} {41--64} (\bibinfo {year} {2019})},\
  \Eprint {http://arxiv.org/abs/1908.02350} {arXiv:1908.02350 [astro-ph.HE]}
  \BibitemShut {NoStop}%
\bibitem [{\citenamefont {{Radice}}\ \emph {et~al.}(2020)\citenamefont
  {{Radice}}, \citenamefont {{Bernuzzi}},\ and\ \citenamefont
  {{Perego}}}]{2020ARNPS..70...95R}%
  \BibitemOpen
  \bibfield  {author} {\bibinfo {author} {\bibfnamefont {David}\ \bibnamefont
  {{Radice}}}, \bibinfo {author} {\bibfnamefont {Sebastiano}\ \bibnamefont
  {{Bernuzzi}}}, \ and\ \bibinfo {author} {\bibfnamefont {Albino}\ \bibnamefont
  {{Perego}}},\ }\bibfield  {title} {\enquote {\bibinfo {title} {{The Dynamics
  of Binary Neutron Star Mergers and GW170817}},}\ }\href {\doibase
  10.1146/annurev-nucl-013120-114541} {\bibfield  {journal} {\bibinfo
  {journal} {Annual Review of Nuclear and Particle Science}\ }\textbf {\bibinfo
  {volume} {70}},\ \bibinfo {pages} {95--119} (\bibinfo {year} {2020})},\
  \Eprint {http://arxiv.org/abs/2002.03863} {arXiv:2002.03863 [astro-ph.HE]}
  \BibitemShut {NoStop}%
\bibitem [{\citenamefont {{Margutti}}\ and\ \citenamefont
  {{Chornock}}(2021)}]{2021ARA&A..59..155M}%
  \BibitemOpen
  \bibfield  {author} {\bibinfo {author} {\bibfnamefont {Raffaella}\
  \bibnamefont {{Margutti}}}\ and\ \bibinfo {author} {\bibfnamefont {Ryan}\
  \bibnamefont {{Chornock}}},\ }\bibfield  {title} {\enquote {\bibinfo {title}
  {{First Multimessenger Observations of a Neutron Star Merger}},}\ }\href
  {\doibase 10.1146/annurev-astro-112420-030742} {\bibfield  {journal}
  {\bibinfo  {journal} {\araa}\ }\textbf {\bibinfo {volume} {59}},\ \bibinfo
  {pages} {155--202} (\bibinfo {year} {2021})},\ \Eprint
  {http://arxiv.org/abs/2012.04810} {arXiv:2012.04810 [astro-ph.HE]}
  \BibitemShut {NoStop}%
\bibitem [{\citenamefont {{Zhu}}\ \emph {et~al.}(2016)\citenamefont {{Zhu}},
  \citenamefont {{Perego}},\ and\ \citenamefont
  {{McLaughlin}}}]{2016PhRvD..94j5006Z}%
  \BibitemOpen
  \bibfield  {author} {\bibinfo {author} {\bibfnamefont {Y.~L.}\ \bibnamefont
  {{Zhu}}}, \bibinfo {author} {\bibfnamefont {A.}~\bibnamefont {{Perego}}}, \
  and\ \bibinfo {author} {\bibfnamefont {G.~C.}\ \bibnamefont {{McLaughlin}}},\
  }\bibfield  {title} {\enquote {\bibinfo {title} {{Matter-neutrino resonance
  transitions above a neutron star merger remnant}},}\ }\href {\doibase
  10.1103/PhysRevD.94.105006} {\bibfield  {journal} {\bibinfo  {journal}
  {\prd}\ }\textbf {\bibinfo {volume} {94}},\ \bibinfo {eid} {105006} (\bibinfo
  {year} {2016})},\ \Eprint {http://arxiv.org/abs/1607.04671} {arXiv:1607.04671
  [hep-ph]} \BibitemShut {NoStop}%
\bibitem [{\citenamefont {{Wu}}\ and\ \citenamefont
  {{Tamborra}}(2017)}]{2017PhRvD..95j3007W}%
  \BibitemOpen
  \bibfield  {author} {\bibinfo {author} {\bibfnamefont {Meng-Ru}\ \bibnamefont
  {{Wu}}}\ and\ \bibinfo {author} {\bibfnamefont {Irene}\ \bibnamefont
  {{Tamborra}}},\ }\bibfield  {title} {\enquote {\bibinfo {title} {{Fast
  neutrino conversions: Ubiquitous in compact binary merger remnants}},}\
  }\href {\doibase 10.1103/PhysRevD.95.103007} {\bibfield  {journal} {\bibinfo
  {journal} {\prd}\ }\textbf {\bibinfo {volume} {95}},\ \bibinfo {eid} {103007}
  (\bibinfo {year} {2017})},\ \Eprint {http://arxiv.org/abs/1701.06580}
  {arXiv:1701.06580 [astro-ph.HE]} \BibitemShut {NoStop}%
\bibitem [{\citenamefont {{Vlasenko}}\ and\ \citenamefont
  {{McLaughlin}}(2018)}]{2018PhRvD..97h3011V}%
  \BibitemOpen
  \bibfield  {author} {\bibinfo {author} {\bibfnamefont {Alexey}\ \bibnamefont
  {{Vlasenko}}}\ and\ \bibinfo {author} {\bibfnamefont {G.~C.}\ \bibnamefont
  {{McLaughlin}}},\ }\bibfield  {title} {\enquote {\bibinfo {title}
  {{Matter-neutrino resonance in a multiangle neutrino bulb model}},}\ }\href
  {\doibase 10.1103/PhysRevD.97.083011} {\bibfield  {journal} {\bibinfo
  {journal} {\prd}\ }\textbf {\bibinfo {volume} {97}},\ \bibinfo {eid} {083011}
  (\bibinfo {year} {2018})},\ \Eprint {http://arxiv.org/abs/1801.07813}
  {arXiv:1801.07813 [astro-ph.HE]} \BibitemShut {NoStop}%
\bibitem [{\citenamefont {{George}}\ \emph {et~al.}(2020)\citenamefont
  {{George}}, \citenamefont {{Wu}}, \citenamefont {{Tamborra}}, \citenamefont
  {{Ardevol-Pulpillo}},\ and\ \citenamefont {{Janka}}}]{2020PhRvD.102j3015G}%
  \BibitemOpen
  \bibfield  {author} {\bibinfo {author} {\bibfnamefont {Manu}\ \bibnamefont
  {{George}}}, \bibinfo {author} {\bibfnamefont {Meng-Ru}\ \bibnamefont
  {{Wu}}}, \bibinfo {author} {\bibfnamefont {Irene}\ \bibnamefont
  {{Tamborra}}}, \bibinfo {author} {\bibfnamefont {Ricard}\ \bibnamefont
  {{Ardevol-Pulpillo}}}, \ and\ \bibinfo {author} {\bibfnamefont {Hans-Thomas}\
  \bibnamefont {{Janka}}},\ }\bibfield  {title} {\enquote {\bibinfo {title}
  {{Fast neutrino flavor conversion, ejecta properties, and nucleosynthesis in
  newly-formed hypermassive remnants of neutron-star mergers}},}\ }\href
  {\doibase 10.1103/PhysRevD.102.103015} {\bibfield  {journal} {\bibinfo
  {journal} {\prd}\ }\textbf {\bibinfo {volume} {102}},\ \bibinfo {eid}
  {103015} (\bibinfo {year} {2020})},\ \Eprint
  {http://arxiv.org/abs/2009.04046} {arXiv:2009.04046 [astro-ph.HE]}
  \BibitemShut {NoStop}%
\bibitem [{\citenamefont {{Li}}\ and\ \citenamefont
  {{Siegel}}(2021)}]{2021PhRvL.126y1101L}%
  \BibitemOpen
  \bibfield  {author} {\bibinfo {author} {\bibfnamefont {Xinyu}\ \bibnamefont
  {{Li}}}\ and\ \bibinfo {author} {\bibfnamefont {Daniel~M.}\ \bibnamefont
  {{Siegel}}},\ }\bibfield  {title} {\enquote {\bibinfo {title} {{Neutrino Fast
  Flavor Conversions in Neutron-Star Postmerger Accretion Disks}},}\ }\href
  {\doibase 10.1103/PhysRevLett.126.251101} {\bibfield  {journal} {\bibinfo
  {journal} {\prl}\ }\textbf {\bibinfo {volume} {126}},\ \bibinfo {eid}
  {251101} (\bibinfo {year} {2021})},\ \Eprint
  {http://arxiv.org/abs/2103.02616} {arXiv:2103.02616 [astro-ph.HE]}
  \BibitemShut {NoStop}%
\bibitem [{\citenamefont {{Richers}}(2022)}]{2022PhRvD.106h3005R}%
  \BibitemOpen
  \bibfield  {author} {\bibinfo {author} {\bibfnamefont {Sherwood}\
  \bibnamefont {{Richers}}},\ }\bibfield  {title} {\enquote {\bibinfo {title}
  {{Evaluating approximate flavor instability metrics in neutron star
  mergers}},}\ }\href {\doibase 10.1103/PhysRevD.106.083005} {\bibfield
  {journal} {\bibinfo  {journal} {\prd}\ }\textbf {\bibinfo {volume} {106}},\
  \bibinfo {eid} {083005} (\bibinfo {year} {2022})},\ \Eprint
  {http://arxiv.org/abs/2206.08444} {arXiv:2206.08444 [astro-ph.HE]}
  \BibitemShut {NoStop}%
\bibitem [{\citenamefont {{Just}}\ \emph {et~al.}(2022)\citenamefont {{Just}},
  \citenamefont {{Abbar}}, \citenamefont {{Wu}}, \citenamefont {{Tamborra}},
  \citenamefont {{Janka}},\ and\ \citenamefont
  {{Capozzi}}}]{2022PhRvD.105h3024J}%
  \BibitemOpen
  \bibfield  {author} {\bibinfo {author} {\bibfnamefont {Oliver}\ \bibnamefont
  {{Just}}}, \bibinfo {author} {\bibfnamefont {Sajad}\ \bibnamefont {{Abbar}}},
  \bibinfo {author} {\bibfnamefont {Meng-Ru}\ \bibnamefont {{Wu}}}, \bibinfo
  {author} {\bibfnamefont {Irene}\ \bibnamefont {{Tamborra}}}, \bibinfo
  {author} {\bibfnamefont {Hans-Thomas}\ \bibnamefont {{Janka}}}, \ and\
  \bibinfo {author} {\bibfnamefont {Francesco}\ \bibnamefont {{Capozzi}}},\
  }\bibfield  {title} {\enquote {\bibinfo {title} {{Fast neutrino conversion in
  hydrodynamic simulations of neutrino-cooled accretion disks}},}\ }\href
  {\doibase 10.1103/PhysRevD.105.083024} {\bibfield  {journal} {\bibinfo
  {journal} {\prd}\ }\textbf {\bibinfo {volume} {105}},\ \bibinfo {eid}
  {083024} (\bibinfo {year} {2022})},\ \Eprint
  {http://arxiv.org/abs/2203.16559} {arXiv:2203.16559 [astro-ph.HE]}
  \BibitemShut {NoStop}%
\bibitem [{\citenamefont {{Grohs}}\ \emph {et~al.}(2022)\citenamefont
  {{Grohs}}, \citenamefont {{Richers}}, \citenamefont {{Couch}}, \citenamefont
  {{Foucart}}, \citenamefont {{Kneller}},\ and\ \citenamefont
  {{McLaughlin}}}]{2022arXiv220702214G}%
  \BibitemOpen
  \bibfield  {author} {\bibinfo {author} {\bibfnamefont {Evan}\ \bibnamefont
  {{Grohs}}}, \bibinfo {author} {\bibfnamefont {Sherwood}\ \bibnamefont
  {{Richers}}}, \bibinfo {author} {\bibfnamefont {Sean~M.}\ \bibnamefont
  {{Couch}}}, \bibinfo {author} {\bibfnamefont {Francois}\ \bibnamefont
  {{Foucart}}}, \bibinfo {author} {\bibfnamefont {James~P.}\ \bibnamefont
  {{Kneller}}}, \ and\ \bibinfo {author} {\bibfnamefont {G.~C.}\ \bibnamefont
  {{McLaughlin}}},\ }\bibfield  {title} {\enquote {\bibinfo {title} {{Neutrino
  Fast Flavor Instability in three dimensions for a Neutron Star Merger}},}\
  }\href@noop {} {\bibfield  {journal} {\bibinfo  {journal} {arXiv e-prints}\
  ,\ \bibinfo {eid} {arXiv:2207.02214}} (\bibinfo {year} {2022})},\ \Eprint
  {http://arxiv.org/abs/2207.02214} {arXiv:2207.02214 [hep-ph]} \BibitemShut
  {NoStop}%
\bibitem [{\citenamefont {{Xiong}}\ \emph {et~al.}(2022)\citenamefont
  {{Xiong}}, \citenamefont {{Johns}}, \citenamefont {{Wu}},\ and\ \citenamefont
  {{Duan}}}]{2022arXiv221203750X}%
  \BibitemOpen
  \bibfield  {author} {\bibinfo {author} {\bibfnamefont {Zewei}\ \bibnamefont
  {{Xiong}}}, \bibinfo {author} {\bibfnamefont {Lucas}\ \bibnamefont
  {{Johns}}}, \bibinfo {author} {\bibfnamefont {Meng-Ru}\ \bibnamefont {{Wu}}},
  \ and\ \bibinfo {author} {\bibfnamefont {Huaiyu}\ \bibnamefont {{Duan}}},\
  }\bibfield  {title} {\enquote {\bibinfo {title} {{Collisional flavor
  instability in dense neutrino gases}},}\ }\href@noop {} {\bibfield  {journal}
  {\bibinfo  {journal} {arXiv e-prints}\ ,\ \bibinfo {eid} {arXiv:2212.03750}}
  (\bibinfo {year} {2022})},\ \Eprint {http://arxiv.org/abs/2212.03750}
  {arXiv:2212.03750 [hep-ph]} \BibitemShut {NoStop}%
\bibitem [{\citenamefont {{Sawyer}}(2005)}]{2005PhRvD..72d5003S}%
  \BibitemOpen
  \bibfield  {author} {\bibinfo {author} {\bibfnamefont {R.~F.}\ \bibnamefont
  {{Sawyer}}},\ }\bibfield  {title} {\enquote {\bibinfo {title} {{Speed-up of
  neutrino transformations in a supernova environment}},}\ }\href {\doibase
  10.1103/PhysRevD.72.045003} {\bibfield  {journal} {\bibinfo  {journal}
  {\prd}\ }\textbf {\bibinfo {volume} {72}},\ \bibinfo {eid} {045003} (\bibinfo
  {year} {2005})},\ \Eprint {http://arxiv.org/abs/hep-ph/0503013}
  {arXiv:hep-ph/0503013 [astro-ph]} \BibitemShut {NoStop}%
\bibitem [{\citenamefont {{Tamborra}}\ and\ \citenamefont
  {{Shalgar}}(2021)}]{2021ARNPS..71..165T}%
  \BibitemOpen
  \bibfield  {author} {\bibinfo {author} {\bibfnamefont {Irene}\ \bibnamefont
  {{Tamborra}}}\ and\ \bibinfo {author} {\bibfnamefont {Shashank}\ \bibnamefont
  {{Shalgar}}},\ }\bibfield  {title} {\enquote {\bibinfo {title} {{New
  Developments in Flavor Evolution of a Dense Neutrino Gas}},}\ }\href
  {\doibase 10.1146/annurev-nucl-102920-050505} {\bibfield  {journal} {\bibinfo
   {journal} {Annual Review of Nuclear and Particle Science}\ }\textbf
  {\bibinfo {volume} {71}},\ \bibinfo {pages} {165--188} (\bibinfo {year}
  {2021})},\ \Eprint {http://arxiv.org/abs/2011.01948} {arXiv:2011.01948
  [astro-ph.HE]} \BibitemShut {NoStop}%
\bibitem [{\citenamefont {{Capozzi}}\ and\ \citenamefont
  {{Saviano}}(2022)}]{2022Univ....8...94C}%
  \BibitemOpen
  \bibfield  {author} {\bibinfo {author} {\bibfnamefont {Francesco}\
  \bibnamefont {{Capozzi}}}\ and\ \bibinfo {author} {\bibfnamefont {Ninetta}\
  \bibnamefont {{Saviano}}},\ }\bibfield  {title} {\enquote {\bibinfo {title}
  {{Neutrino Flavor Conversions in High-Density Astrophysical and Cosmological
  Environments}},}\ }\href {\doibase 10.3390/universe8020094} {\bibfield
  {journal} {\bibinfo  {journal} {Universe}\ }\textbf {\bibinfo {volume} {8}},\
  \bibinfo {pages} {94} (\bibinfo {year} {2022})},\ \Eprint
  {http://arxiv.org/abs/2202.02494} {arXiv:2202.02494 [hep-ph]} \BibitemShut
  {NoStop}%
\bibitem [{\citenamefont {{Richers}}\ and\ \citenamefont
  {{Sen}}(2022)}]{2022arXiv220703561R}%
  \BibitemOpen
  \bibfield  {author} {\bibinfo {author} {\bibfnamefont {Sherwood}\
  \bibnamefont {{Richers}}}\ and\ \bibinfo {author} {\bibfnamefont {Manibrata}\
  \bibnamefont {{Sen}}},\ }\bibfield  {title} {\enquote {\bibinfo {title}
  {{Fast Flavor Transformations}},}\ }\href@noop {} {\bibfield  {journal}
  {\bibinfo  {journal} {arXiv e-prints}\ ,\ \bibinfo {eid} {arXiv:2207.03561}}
  (\bibinfo {year} {2022})},\ \Eprint {http://arxiv.org/abs/2207.03561}
  {arXiv:2207.03561 [astro-ph.HE]} \BibitemShut {NoStop}%
\bibitem [{\citenamefont {{Volpe}}(2023)}]{2023arXiv230111814V}%
  \BibitemOpen
  \bibfield  {author} {\bibinfo {author} {\bibfnamefont {Maria~Cristina}\
  \bibnamefont {{Volpe}}},\ }\bibfield  {title} {\enquote {\bibinfo {title}
  {{Neutrinos from dense: flavor mechanisms, theoretical approaches,
  observations, new directions}},}\ }\href {\doibase 10.48550/arXiv.2301.11814}
  {\bibfield  {journal} {\bibinfo  {journal} {arXiv e-prints}\ ,\ \bibinfo
  {eid} {arXiv:2301.11814}} (\bibinfo {year} {2023})},\ \Eprint
  {http://arxiv.org/abs/2301.11814} {arXiv:2301.11814 [hep-ph]} \BibitemShut
  {NoStop}%
\bibitem [{\citenamefont {{Padilla-Gay}}\ \emph {et~al.}(2021)\citenamefont
  {{Padilla-Gay}}, \citenamefont {{Shalgar}},\ and\ \citenamefont
  {{Tamborra}}}]{2021JCAP...01..017P}%
  \BibitemOpen
  \bibfield  {author} {\bibinfo {author} {\bibfnamefont {Ian}\ \bibnamefont
  {{Padilla-Gay}}}, \bibinfo {author} {\bibfnamefont {Shashank}\ \bibnamefont
  {{Shalgar}}}, \ and\ \bibinfo {author} {\bibfnamefont {Irene}\ \bibnamefont
  {{Tamborra}}},\ }\bibfield  {title} {\enquote {\bibinfo {title}
  {{Multi-dimensional solution of fast neutrino conversions in binary neutron
  star merger remnants}},}\ }\href {\doibase 10.1088/1475-7516/2021/01/017}
  {\bibfield  {journal} {\bibinfo  {journal} {\jcap}\ }\textbf {\bibinfo
  {volume} {2021}},\ \bibinfo {eid} {017} (\bibinfo {year} {2021})},\ \Eprint
  {http://arxiv.org/abs/2009.01843} {arXiv:2009.01843 [astro-ph.HE]}
  \BibitemShut {NoStop}%
\bibitem [{\citenamefont {{Fern{\'a}ndez}}\ \emph {et~al.}(2022)\citenamefont
  {{Fern{\'a}ndez}}, \citenamefont {{Richers}}, \citenamefont {{Mulyk}},\ and\
  \citenamefont {{Fahlman}}}]{2022PhRvD.106j3003F}%
  \BibitemOpen
  \bibfield  {author} {\bibinfo {author} {\bibfnamefont {Rodrigo}\ \bibnamefont
  {{Fern{\'a}ndez}}}, \bibinfo {author} {\bibfnamefont {Sherwood}\ \bibnamefont
  {{Richers}}}, \bibinfo {author} {\bibfnamefont {Nicole}\ \bibnamefont
  {{Mulyk}}}, \ and\ \bibinfo {author} {\bibfnamefont {Steven}\ \bibnamefont
  {{Fahlman}}},\ }\bibfield  {title} {\enquote {\bibinfo {title} {{Fast flavor
  instability in hypermassive neutron star disk outflows}},}\ }\href {\doibase
  10.1103/PhysRevD.106.103003} {\bibfield  {journal} {\bibinfo  {journal}
  {\prd}\ }\textbf {\bibinfo {volume} {106}},\ \bibinfo {eid} {103003}
  (\bibinfo {year} {2022})},\ \Eprint {http://arxiv.org/abs/2207.10680}
  {arXiv:2207.10680 [astro-ph.HE]} \BibitemShut {NoStop}%
\bibitem [{\citenamefont {{Nagakura}}\ and\ \citenamefont
  {{Zaizen}}(2022)}]{2022PhRvL.129z1101N}%
  \BibitemOpen
  \bibfield  {author} {\bibinfo {author} {\bibfnamefont {Hiroki}\ \bibnamefont
  {{Nagakura}}}\ and\ \bibinfo {author} {\bibfnamefont {Masamichi}\
  \bibnamefont {{Zaizen}}},\ }\bibfield  {title} {\enquote {\bibinfo {title}
  {{Time-Dependent and Quasisteady Features of Fast Neutrino-Flavor
  Conversion}},}\ }\href {\doibase 10.1103/PhysRevLett.129.261101} {\bibfield
  {journal} {\bibinfo  {journal} {\prl}\ }\textbf {\bibinfo {volume} {129}},\
  \bibinfo {eid} {261101} (\bibinfo {year} {2022})},\ \Eprint
  {http://arxiv.org/abs/2206.04097} {arXiv:2206.04097 [astro-ph.HE]}
  \BibitemShut {NoStop}%
\bibitem [{\citenamefont {{Nagakura}}\ and\ \citenamefont
  {{Zaizen}}(2023{\natexlab{a}})}]{2023PhRvD.107f3033N}%
  \BibitemOpen
  \bibfield  {author} {\bibinfo {author} {\bibfnamefont {Hiroki}\ \bibnamefont
  {{Nagakura}}}\ and\ \bibinfo {author} {\bibfnamefont {Masamichi}\
  \bibnamefont {{Zaizen}}},\ }\bibfield  {title} {\enquote {\bibinfo {title}
  {{Connecting small-scale to large-scale structures of fast neutrino-flavor
  conversion}},}\ }\href {\doibase 10.1103/PhysRevD.107.063033} {\bibfield
  {journal} {\bibinfo  {journal} {\prd}\ }\textbf {\bibinfo {volume} {107}},\
  \bibinfo {eid} {063033} (\bibinfo {year} {2023}{\natexlab{a}})},\ \Eprint
  {http://arxiv.org/abs/2211.01398} {arXiv:2211.01398 [astro-ph.HE]}
  \BibitemShut {NoStop}%
\bibitem [{\citenamefont {{Nagakura}}(2023)}]{2023PhRvL.130u1401N}%
  \BibitemOpen
  \bibfield  {author} {\bibinfo {author} {\bibfnamefont {Hiroki}\ \bibnamefont
  {{Nagakura}}},\ }\bibfield  {title} {\enquote {\bibinfo {title} {{Roles of
  Fast Neutrino-Flavor Conversion on the Neutrino-Heating Mechanism of
  Core-Collapse Supernova}},}\ }\href {\doibase 10.1103/PhysRevLett.130.211401}
  {\bibfield  {journal} {\bibinfo  {journal} {\prl}\ }\textbf {\bibinfo
  {volume} {130}},\ \bibinfo {eid} {211401} (\bibinfo {year} {2023})},\ \Eprint
  {http://arxiv.org/abs/2301.10785} {arXiv:2301.10785 [astro-ph.HE]}
  \BibitemShut {NoStop}%
\bibitem [{\citenamefont {{Johns}}\ and\ \citenamefont
  {{Xiong}}(2022)}]{2022PhRvD.106j3029J}%
  \BibitemOpen
  \bibfield  {author} {\bibinfo {author} {\bibfnamefont {Lucas}\ \bibnamefont
  {{Johns}}}\ and\ \bibinfo {author} {\bibfnamefont {Zewei}\ \bibnamefont
  {{Xiong}}},\ }\bibfield  {title} {\enquote {\bibinfo {title} {{Collisional
  instabilities of neutrinos and their interplay with fast flavor conversion in
  compact objects}},}\ }\href {\doibase 10.1103/PhysRevD.106.103029} {\bibfield
   {journal} {\bibinfo  {journal} {\prd}\ }\textbf {\bibinfo {volume} {106}},\
  \bibinfo {eid} {103029} (\bibinfo {year} {2022})},\ \Eprint
  {http://arxiv.org/abs/2208.11059} {arXiv:2208.11059 [hep-ph]} \BibitemShut
  {NoStop}%
\bibitem [{\citenamefont {{Kato}}\ and\ \citenamefont
  {{Nagakura}}(2022)}]{2022PhRvD.106l3013K}%
  \BibitemOpen
  \bibfield  {author} {\bibinfo {author} {\bibfnamefont {Chinami}\ \bibnamefont
  {{Kato}}}\ and\ \bibinfo {author} {\bibfnamefont {Hiroki}\ \bibnamefont
  {{Nagakura}}},\ }\bibfield  {title} {\enquote {\bibinfo {title} {{Effects of
  energy-dependent scatterings on fast neutrino flavor conversions}},}\ }\href
  {\doibase 10.1103/PhysRevD.106.123013} {\bibfield  {journal} {\bibinfo
  {journal} {\prd}\ }\textbf {\bibinfo {volume} {106}},\ \bibinfo {eid}
  {123013} (\bibinfo {year} {2022})},\ \Eprint
  {http://arxiv.org/abs/2207.09496} {arXiv:2207.09496 [astro-ph.HE]}
  \BibitemShut {NoStop}%
\bibitem [{\citenamefont {{Kato}}\ \emph {et~al.}(2023)\citenamefont {{Kato}},
  \citenamefont {{Nagakura}},\ and\ \citenamefont
  {{Zaizen}}}]{2023arXiv230316453K}%
  \BibitemOpen
  \bibfield  {author} {\bibinfo {author} {\bibfnamefont {Chinami}\ \bibnamefont
  {{Kato}}}, \bibinfo {author} {\bibfnamefont {Hiroki}\ \bibnamefont
  {{Nagakura}}}, \ and\ \bibinfo {author} {\bibfnamefont {Masamichi}\
  \bibnamefont {{Zaizen}}},\ }\bibfield  {title} {\enquote {\bibinfo {title}
  {{Flavor conversions with energy-dependent neutrino emission and
  absorption}},}\ }\href {\doibase 10.1103/PhysRevD.108.023006} {\bibfield
  {journal} {\bibinfo  {journal} {\prd}\ }\textbf {\bibinfo {volume} {108}},\
  \bibinfo {eid} {023006} (\bibinfo {year} {2023})},\ \Eprint
  {http://arxiv.org/abs/2303.16453} {arXiv:2303.16453 [astro-ph.HE]}
  \BibitemShut {NoStop}%
\bibitem [{\citenamefont {{Nagakura}}(2022)}]{2022PhRvD.106f3011N}%
  \BibitemOpen
  \bibfield  {author} {\bibinfo {author} {\bibfnamefont {Hiroki}\ \bibnamefont
  {{Nagakura}}},\ }\bibfield  {title} {\enquote {\bibinfo {title}
  {{General-relativistic quantum-kinetics neutrino transport}},}\ }\href
  {\doibase 10.1103/PhysRevD.106.063011} {\bibfield  {journal} {\bibinfo
  {journal} {\prd}\ }\textbf {\bibinfo {volume} {106}},\ \bibinfo {eid}
  {063011} (\bibinfo {year} {2022})},\ \Eprint
  {http://arxiv.org/abs/2206.04098} {arXiv:2206.04098 [astro-ph.HE]}
  \BibitemShut {NoStop}%
\bibitem [{\citenamefont {{Sumiyoshi}}\ \emph {et~al.}(2021)\citenamefont
  {{Sumiyoshi}}, \citenamefont {{Fujibayashi}}, \citenamefont {{Sekiguchi}},\
  and\ \citenamefont {{Shibata}}}]{2021ApJ...907...92S}%
  \BibitemOpen
  \bibfield  {author} {\bibinfo {author} {\bibfnamefont {Kohsuke}\ \bibnamefont
  {{Sumiyoshi}}}, \bibinfo {author} {\bibfnamefont {Sho}\ \bibnamefont
  {{Fujibayashi}}}, \bibinfo {author} {\bibfnamefont {Yuichiro}\ \bibnamefont
  {{Sekiguchi}}}, \ and\ \bibinfo {author} {\bibfnamefont {Masaru}\
  \bibnamefont {{Shibata}}},\ }\bibfield  {title} {\enquote {\bibinfo {title}
  {{Properties of Neutrino Transfer in a Deformed Remnant of a Neutron Star
  Merger}},}\ }\href {\doibase 10.3847/1538-4357/abce63} {\bibfield  {journal}
  {\bibinfo  {journal} {\apj}\ }\textbf {\bibinfo {volume} {907}},\ \bibinfo
  {eid} {92} (\bibinfo {year} {2021})},\ \Eprint
  {http://arxiv.org/abs/2010.10865} {arXiv:2010.10865 [astro-ph.HE]}
  \BibitemShut {NoStop}%
\bibitem [{\citenamefont {{Zaizen}}\ and\ \citenamefont
  {{Nagakura}}(2023)}]{2023arXiv230405044Z}%
  \BibitemOpen
  \bibfield  {author} {\bibinfo {author} {\bibfnamefont {Masamichi}\
  \bibnamefont {{Zaizen}}}\ and\ \bibinfo {author} {\bibfnamefont {Hiroki}\
  \bibnamefont {{Nagakura}}},\ }\bibfield  {title} {\enquote {\bibinfo {title}
  {{Characterizing quasisteady states of fast neutrino-flavor conversion by
  stability and conservation laws}},}\ }\href {\doibase
  10.1103/PhysRevD.107.123021} {\bibfield  {journal} {\bibinfo  {journal}
  {\prd}\ }\textbf {\bibinfo {volume} {107}},\ \bibinfo {eid} {123021}
  (\bibinfo {year} {2023})},\ \Eprint {http://arxiv.org/abs/2304.05044}
  {arXiv:2304.05044 [astro-ph.HE]} \BibitemShut {NoStop}%
\bibitem [{\citenamefont {{Dasgupta}}\ \emph {et~al.}(2008)\citenamefont
  {{Dasgupta}}, \citenamefont {{Dighe}}, \citenamefont {{Mirizzi}},\ and\
  \citenamefont {{Raffelt}}}]{2008PhRvD..78c3014D}%
  \BibitemOpen
  \bibfield  {author} {\bibinfo {author} {\bibfnamefont {Basudeb}\ \bibnamefont
  {{Dasgupta}}}, \bibinfo {author} {\bibfnamefont {Amol}\ \bibnamefont
  {{Dighe}}}, \bibinfo {author} {\bibfnamefont {Alessandro}\ \bibnamefont
  {{Mirizzi}}}, \ and\ \bibinfo {author} {\bibfnamefont {Georg}\ \bibnamefont
  {{Raffelt}}},\ }\bibfield  {title} {\enquote {\bibinfo {title} {{Collective
  neutrino oscillations in nonspherical geometry}},}\ }\href {\doibase
  10.1103/PhysRevD.78.033014} {\bibfield  {journal} {\bibinfo  {journal}
  {\prd}\ }\textbf {\bibinfo {volume} {78}},\ \bibinfo {eid} {033014} (\bibinfo
  {year} {2008})},\ \Eprint {http://arxiv.org/abs/0805.3300} {arXiv:0805.3300
  [hep-ph]} \BibitemShut {NoStop}%
\bibitem [{\citenamefont {{Nagakura}}\ and\ \citenamefont
  {{Zaizen}}(2023{\natexlab{b}})}]{2022arXiv221101398N}%
  \BibitemOpen
  \bibfield  {author} {\bibinfo {author} {\bibfnamefont {Hiroki}\ \bibnamefont
  {{Nagakura}}}\ and\ \bibinfo {author} {\bibfnamefont {Masamichi}\
  \bibnamefont {{Zaizen}}},\ }\bibfield  {title} {\enquote {\bibinfo {title}
  {{Connecting small-scale to large-scale structures of fast neutrino-flavor
  conversion}},}\ }\href {\doibase 10.1103/PhysRevD.107.063033} {\bibfield
  {journal} {\bibinfo  {journal} {\prd}\ }\textbf {\bibinfo {volume} {107}},\
  \bibinfo {eid} {063033} (\bibinfo {year} {2023}{\natexlab{b}})},\ \Eprint
  {http://arxiv.org/abs/2211.01398} {arXiv:2211.01398 [astro-ph.HE]}
  \BibitemShut {NoStop}%
\end{thebibliography}%

\end{document}